\documentclass[12pt]{article}

\usepackage{graphicx,amsmath,amssymb,url}

\newlength{\dinwidth}
\newlength{\dinmargin}
\def\lapproxeq{\lower .7ex\hbox{$\;\stackrel{\textstyle
<}{\sim}\;$}}
\def\gapproxeq{\lower .7ex\hbox{$\;\stackrel{\textstyle
>}{\sim}\;$}}
\def\gtrsim{\lower .7ex\hbox{$\;\stackrel{\textstyle
>}{\sim}\;$}}
\def\lesim{\lower .7ex\hbox{$\;\stackrel{\textstyle
<}{\sim}\;$}}

\def\be{\begin{equation}}
\def\ee{\end{equation}}
\def\bea{\begin{eqnarray}}
\def\eea{\end{eqnarray}}

\def\GeV{\rm GeV}
\def\J{J/\psi}

\def\psitwos{\psi(2{\rm S})}
\def\psiprime{\psi(2{\rm S})} 

\begin{document}

\titlepage

\begin{flushright}

IPPP/13/107\\

DCPT/13/214\\

LTH 996\\

24 February 2014\\

\end{flushright}

\vspace*{2cm}

\begin{center}

{\Large \bf Predictions of exclusive $\psitwos$ production at the LHC}

\vspace*{1cm} {\sc S.P. Jones}$^a$, {\sc A.D. Martin}$^b$,  {\sc
  M.G. Ryskin}$^{b, c}$ and {\sc T. Teubner}$^{a}$ \\

\vspace*{0.5cm}
$^a$ {\em Department of Mathematical Sciences,\\

 University of Liverpool, Liverpool L69 3BX, U.K.}\\
$^b$ {\em Department of Physics and Institute for Particle Physics
  Phenomenology,\\ 

University of Durham, Durham DH1 3LE, U.K.}\\

$^c$ {\em Petersburg Nuclear Physics Institute, NRC Kurchatov Institute, Gatchina,
St.~Petersburg, 188300, Russia} \\

 \end{center}

\vspace*{1cm}

\begin{abstract}
The cross section for exclusive $\psitwos$ ultraperipheral production
at the LHC is calculated using gluon parametrisations extracted from
exclusive $\J$ measurements performed at HERA and the LHC.
Predictions are given at leading and next-to-leading order for $pp$ 
centre-of-mass energies of $7$, $8$ and $14$ TeV, assuming the 
non-relativistic approximation for the $\psitwos$ wave function. 
\end{abstract}

\vspace*{0.5cm}

Recently, measurements of exclusive $\J$ production in
ultraperipheral $pp$ and $Pb$-$Pb$ collision have been published by
the LHCb and ALICE collaborations,
\cite{Aaij:2013jxj,Abelev:2012ba}. More data,
with better statistics for the $\J$ and the $\psitwos$ than the
published results, are currently being analysed.\footnote{
Note added in proof: these data have subsequently been published 
\cite{Aaij:2014iea}}
In this short note, using the framework and gluon parametrisations
from~\cite{jmrt}, we make predictions for the exclusive $\psitwos$
production at the LHC for $pp$ centre-of-mass energies of $\sqrt{s}=
7$, $8$ and $14$ TeV.  Alternative predictions for exclusive
$\psitwos$ production at the LHC within the dipole formalism are given
in Refs.~\cite{Ducati:2013tva,Ducati:2013bya}.

For clarity, let us repeat the main formulae used for the leading
order (LO) and next-to-leading order (NLO) predictions. 
The LO photoproduction cross section for $\gamma p \to \psitwos\,p$
is driven by the gluon distribution $xg$ and, in the case of zero $t$-channel 
momentum transfer ($t=0$), is given by~\cite{Ryskin:1992ui} 
\begin{equation}
\frac{{\rm d}\sigma}{{\rm d}t}\left( \gamma p \to \psiprime ~p \right)
     {\Big |}_{t=0} = \frac{\Gamma_{ee}M^3_{\psiprime}\pi^3}{48\alpha}\,
     \left[\frac{{\alpha_s(\bar Q^2)}}{\bar Q^4}xg(x,\bar
     Q^2)\right]^2\,,
\label{eq:lo}
\end{equation}
where $\alpha$ is the QED coupling, $M_{\psiprime}$ is the mass of the $\psitwos$ and $\Gamma_{ee}$
is its electronic width. 
For photoproduction the kinematic variables are 
\begin{equation}
{\bar Q^2}~=~M^2_{\psiprime}/4\,, ~~~~~~~~~x~=~M^2_{\psiprime}/W^2\,,
\end{equation}
and $W$ is the $\gamma p$ centre-of-mass energy. In order to include data
integrated over $t$ we assume the cross section depends exponentially on $t$, 
i.e. $\sigma \sim \exp(-B|t|)$. The energy-dependent $t$ slope parameter, $B$, is, 
given by the Regge motivated form
\begin{equation}
B(W) = \left(4.9 + 4 \alpha' \ln(W/W_0)\right) {\rm\
  GeV}^{-2}\,,
\label{eq:b-slope}
\end{equation}
where the pomeron slope $\alpha'=0.06$ and $W_0=90$~GeV, essentially
unchanged from the $\J$ case. Corrections
due to the skewing of the gluons and the real part of the amplitude
are included as in~\cite{jmrt}. 

At NLO we account for the fact that no additional gluons with
transverse momentum larger than $k_T$ are emitted in the process
by including the Sudakov factor
\begin{equation}
T(k_T^2,\mu^2)={\rm exp}\left[\frac{-C_A\alpha_s(\mu^2)}{4\pi}{\rm ln}^2\left(\frac{\mu^2}{k_T^2}\right)\right]
\end{equation}
with $T=1$ for $k^2_T \geq \mu^2$. Integrating over the $k_T$ of the
gluons, the `NLO' cross section\footnote{By integrating over the gluon
$k_T$ we account for an important part of the next-to-leading order
effects, although we do not include the full set of NLO corrections
to the hard matrix element.} is obtained, as derived in~\cite{jmrt},
by the replacement 
\begin{align}
\left[\frac{{\alpha_s(\bar Q^2)}}{\bar Q^4}xg(x,\bar Q^2)\right]
\:\longrightarrow\: &
\int_{Q_0^2}^{(W^2-M_{\psiprime}^2)/4} 
\frac{{\rm d}k_T^2\,\alpha_s(\mu^2)}{\bar Q^2 (\bar Q^2 + k_T^2)} \, 
\frac{\partial \left[ xg(x,k_T^2) \sqrt{T(k^2_T,\mu^2)}
  \right]}{\partial k_T^2}  \nonumber \\ 
& +\ \ln\left( \frac{\bar{Q}^2+Q^2_0}{\bar{Q}^2}\right)
\frac{\alpha_s(\mu^2_{\rm IR})}{\bar{Q}^2 Q^2_0} \, xg(x,Q_0^2) 
\sqrt{T(Q^2_0,\mu^2_{\rm IR})}\,.
\label{eq:nlointegral}
\end{align}
Here we have assumed the behaviour of $xg(x,k_T^2)\sqrt{T}$ to be
linear in $k_T^2$ for $k_T$ below the infra-red scale $Q_0=1$ GeV. The
scales are chosen to be $\mu^2=\max(k_T^2,\bar{Q}^2)$ and $\mu^2_{\rm
  IR}=\max(Q_0^2,\bar{Q}^2)$. When evaluating (\ref{eq:lo}) and
(\ref{eq:nlointegral}) we use the LO and NLO gluon parametrisations
fitted in~\cite{jmrt}.

\begin{figure}
\begin{center}
\includegraphics[width=0.5\textwidth,angle=-90]{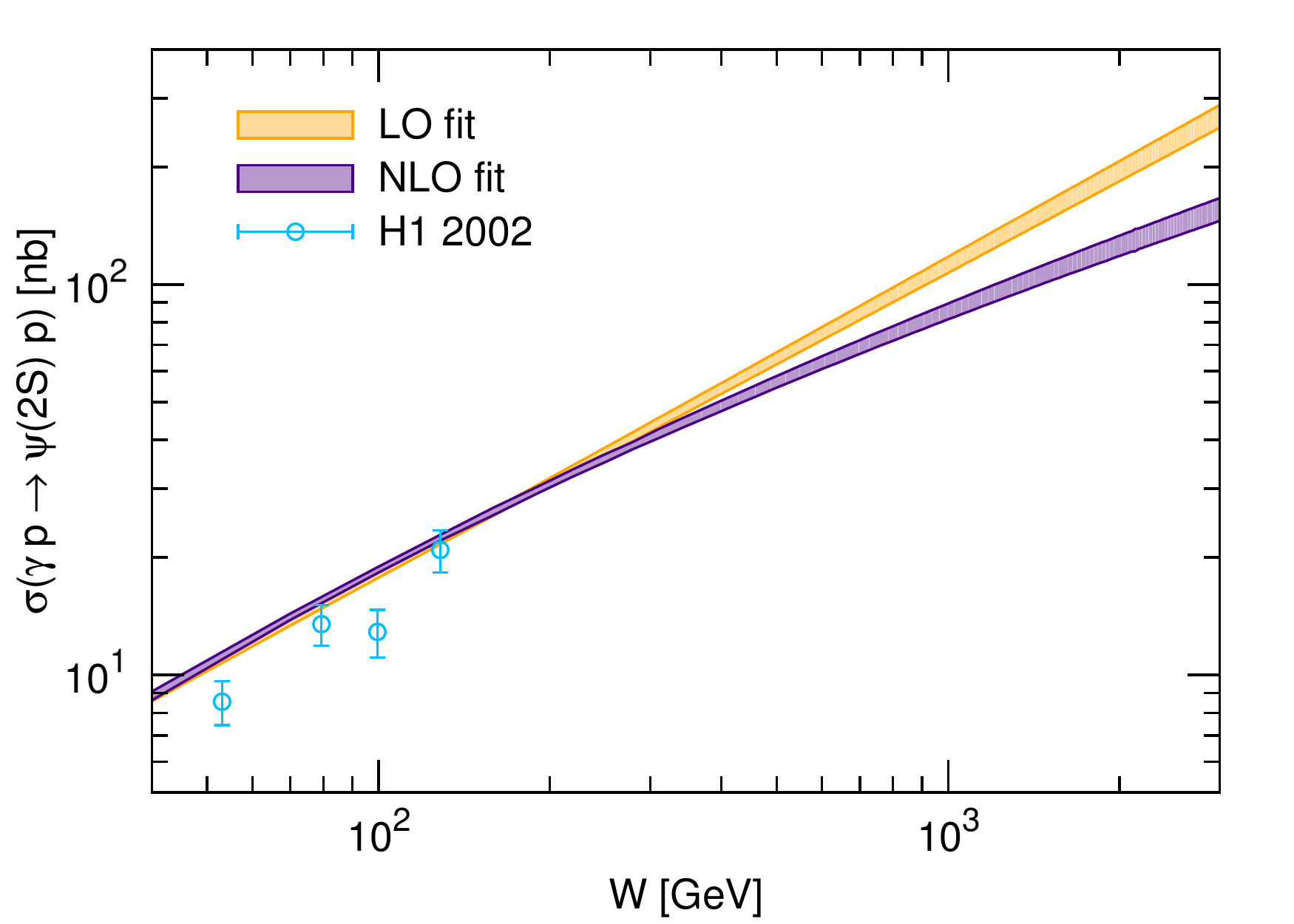}
\caption{Prediction of exclusive $\psitwos$ photoproduction as a
  function of the $\gamma p$ centre-of-mass energy. Also shown,
  but not fitted, are the available $\psitwos$ data from H1
  \cite{Adloff:2002re,Alexa:2013xxa}. The width of the shaded bands
  indicates only the $1\sigma$ uncertainty from the $\J$ experimental
  data used in the gluon fits.} 
\label{fig:sigmapsi2sw}
\end{center}
\end{figure}

Note that for $\psitwos$ the relativistic corrections due to the
vector meson wave function may be larger than for the $\J$, where
they were found to suppress the cross section by about 6\% \cite{Hoodbhoy:1996zg}. 
While a part of the relativistic corrections
related to the wave function at the origin is accounted for using 
$\Gamma_{ee}$, the measured electronic width of the $\psitwos$, one may
expect a further suppression in the case of $\psitwos$ compared to
$\J$. We do not account for this. 

What additional uncertainties are there in our predictions for the gluon
distribution? First, there may be an uncertainty arising from the skewed factor 
which accounts for the difference between  the conventional (diagonal) gluon 
PDF and the generalised (GPD) distribution.
As described in~\cite{jmrt}, we use the Shuvaev transform to relate the
diagonal PDF and the GPD. This provides sufficient accuracy,
$\sim \mathcal{O}(x)$, in our low $x$ domain. 
Next, there may be an uncertainty coming from the real part of the amplitude, 
which is evaluated approximately. 
Again the corresponding uncertainty is small (less than $2\%$)
in the low $x$ region, where the $x$ dependence is not steep and the 
${\rm Re}/{\rm Im}$ ratio is rather small. We emphasize that the uncertainties,
both from the skewed factor and the ${\rm Re}/{\rm Im}$ ratio, apply to $xg(x)$ and {\it not} to
the ratio of the $\psitwos$ to $\J$ cross sections in which they cancel 
almost exactly. Thus, the main uncertainty in the $\psitwos$ cross section, 
calculated using the gluon extracted from the $\J$ data, is that coming 
from the relativistic correction for $\psitwos$.

Figure~\ref{fig:sigmapsi2sw} displays our results for the $\gamma p
\to \psitwos\,p$ cross section in LO and NLO. The width of the shaded bands 
gives the $1\sigma$ uncertainty from the $\J$ experimental data used 
in the gluon fits. Note that we do not include the available $\psitwos$
data from H1~\cite{Adloff:2002re} in the gluon fit, they are shown just for comparison. 
The data~\cite{Adloff:2002re} only gives the values of the ratio of the $\psitwos$ to $\J$
cross sections. To obtain the $\psitwos$ photoproduction data points displayed in 
Fig.~\ref{fig:sigmapsi2sw} we use the fit $\sigma_{\J}=81~(W/90~{\GeV})^{0.67}$~nb 
for the $\J$ cross section obtained by H1~\cite{Alexa:2013xxa}. 
Our predictions are slightly above the data.

\begin{figure}
\begin{center}
\includegraphics[width=0.8\textwidth,clip=true,trim=0em 15em 0em 10em]{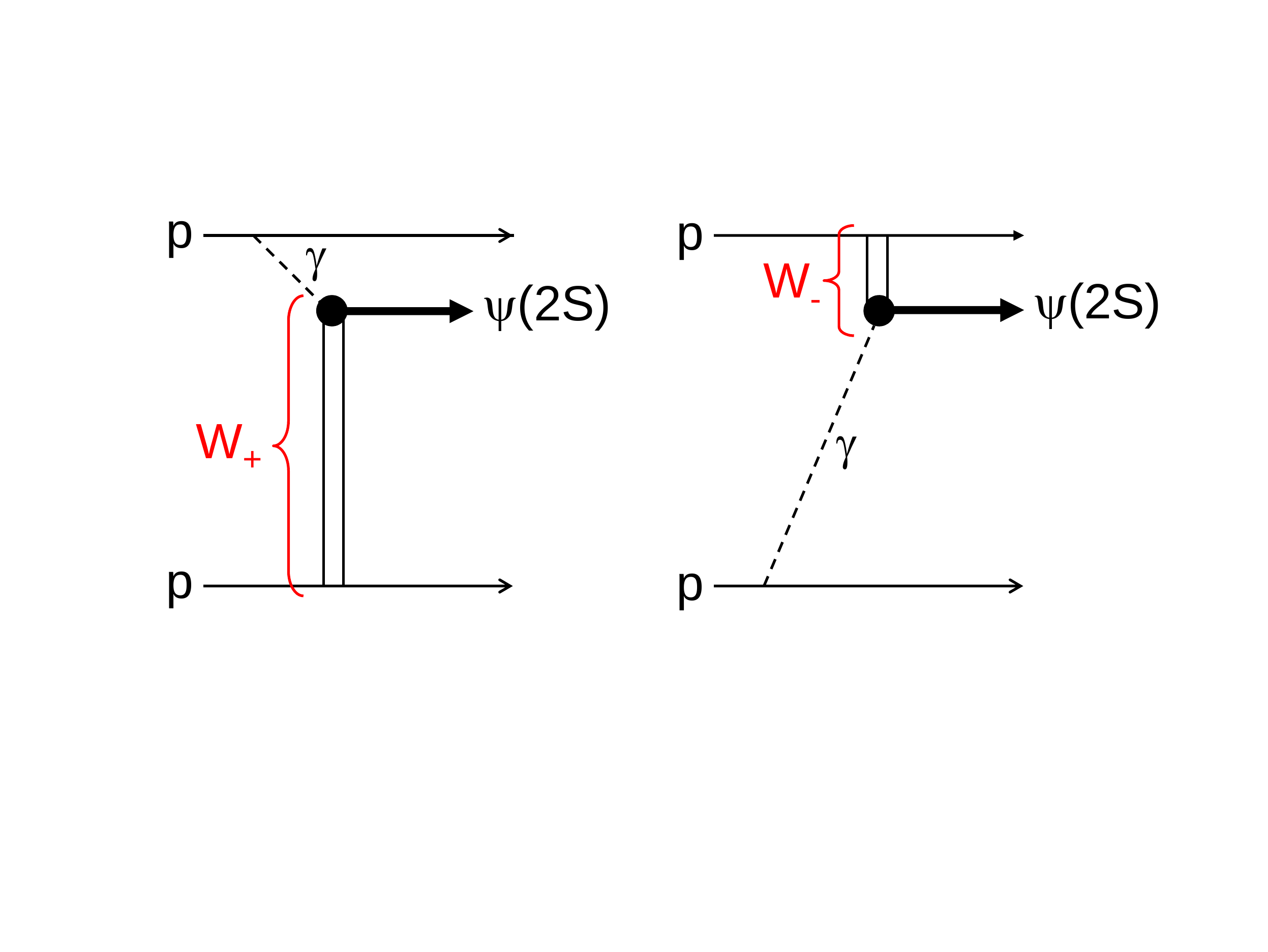}
\vspace{-3mm}
\caption{Subprocesses contributing to exclusive $\psitwos$ production
  in ultraperipheral $pp$ collisions. $W_+$ and $W_-$ are the $\gamma
  p$ centre-of-mass energies. The vertical axis corresponds to the $\psitwos$ rapidity.} 
\label{fig:psi2sproc}
\end{center}
\end{figure}

We can now make predictions for exclusive $\psitwos$ production in 
ultraperipheral $pp$ collisions as a function of the $\psitwos$ rapidity,
using the $\gamma p \to \psitwos\,p$ cross section. Note that
for a given rapidity $y$, two $\gamma p$ subprocesses with different
$\gamma p$ centre-of-mass energies squared, 
$W^2_\pm = M_{\psiprime}\sqrt{s} \exp(\pm|y|)$, and different photon fluxes, 
${\rm d}n/{\rm d}k_\pm$, contribute, depending on which of the protons acts as
photon emitter and which as target, as illustrated in Fig.~\ref{fig:psi2sproc}. In
the absence of forward proton tagging these two subprocesses can not
be distinguished and must be added, see~(\ref{eq:sigmappth}) below. 

Recall that our analysis is well
justified at small $x$, $x \lapproxeq 0.01$, which provides good
accuracy of both the Shuvaev transform, used to relate the diagonal gluons
and GPDs, and the real part contribution. Moreover, the simple expression
used to parametrize the gluon distribution is not appropriate at large $x$.
The $\J$ data included in the gluon fit~\cite{jmrt}, used in this analysis, contains the low energy $W_-$ contribution to the LHC data for which the small $x$ approximation is not justified. In future analyses, it may be better to repeat the gluon fit after subtracting the contribution arising from the low energy $W_-$ configuration from the LHC $J/\psi$ ultraperipheral $pp$ production data. To estimate this contribution a fit to the low energy fixed target data measured at E401 and E516~\cite{Binkley:1981kv,Denby:1983az} could be used. This low energy fixed target fit along with a further fit to the ratio of the $\psitwos$ to $\J$ photoproduction cross sections, measured by H1~\cite{Adloff:2002re}, would allow the  $W_-$ contribution to $\psitwos$ ultraperipheral $pp$ production cross sections to be estimated. Note, however, for $\psitwos$ the size of the $W_-$ contribution is small relative to the $W_+$ contribution due to the lower photon energy and small low energy photoproduction cross section. Thus, though our theory prediction overestimates the $W_-$ contribution to the ultraperipheral $pp$ production its contribution is dominated by the $W_+$ contribution.

\begin{table}[htb]
\begin{center}
\begin{tabular}{|c|c|c|c|c|c|c|}\hline
  &  \multicolumn{2}{|c|} {7 TeV} &  \multicolumn{2}{|c|} {8 TeV} & \multicolumn{2}{c|} {14 TeV}\\   \hline
$y$ &   $S^2(W_+)$ &  $S^2(W_-)$ & $S^2(W_+)$ &  $S^2(W_-)$ & $S^2(W_+)$ &  $S^2(W_-)$   \\ \hline
0.125	&	0.858	&	0.864	&	0.860	&	0.865	&	0.867	&	0.871	\\
0.375	&	0.853	&	0.869	&	0.855	&	0.870	&	0.862	&	0.875	\\
0.625	&	0.846	&	0.873	&	0.848	&	0.875	&	0.856	&	0.879	\\
0.875	&	0.839	&	0.878	&	0.842	&	0.879	&	0.851	&	0.883	\\
1.125	&	0.832	&	0.882	&	0.835	&	0.883	&	0.845	&	0.887	\\
1.375	&	0.824	&	0.885	&	0.827	&	0.886	&	0.838	&	0.890	\\
1.625	&	0.815	&	0.889	&	0.818	&	0.890	&	0.831	&	0.893	\\
1.875	&	0.805	&	0.892	&	0.809	&	0.893	&	0.823	&	0.896	\\
2.125	&	0.794	&	0.895	&	0.798	&	0.896	&	0.814	&	0.899	\\
2.375	&	0.782	&	0.898	&	0.787	&	0.899	&	0.804	&	0.902	\\
2.625	&	0.768	&	0.901	&	0.774	&	0.902	&	0.794	&	0.904	\\
2.875	&	0.753	&	0.904	&	0.759	&	0.904	&	0.782	&	0.906	\\
3.125	&	0.736	&	0.906	&	0.743	&	0.907	&	0.769	&	0.909	\\
3.375	&	0.717	&	0.909	&	0.725	&	0.909	&	0.754	&	0.911	\\
3.625	&	0.696	&	0.911	&	0.705	&	0.911	&	0.738	&	0.913	\\
3.875	&	0.673	&	0.913	&	0.683	&	0.913	&	0.720	&	0.915	\\
4.125	&	0.649	&	0.915	&	0.659	&	0.915	&	0.700	&	0.917	\\
4.375	&	0.624	&	0.917	&	0.634	&	0.917	&	0.677	&	0.919	\\
4.625	&	0.600	&	0.919	&	0.609	&	0.919	&	0.653	&	0.920	\\
4.875	&	0.582	&	0.921	&	0.588	&	0.921	&	0.628	&	0.922	\\
5.125	&	0.573	&	0.922	&	0.573	&	0.923	&	0.602	&	0.924	\\
5.375	&	0.581	&	0.924	&	0.571	&	0.924	&	0.580	&	0.925	\\
5.625	&	0.611	&	0.926	&	0.590	&	0.926	&	0.563	&	0.927	\\
5.875	&	0.666	&	0.927	&	0.632	&	0.927	&	0.559	&	0.928	\\
\hline
\end{tabular}
\end{center}
\caption{Rapidity gap survival factors $S^2$ for exclusive $\psitwos$
  production, $pp \to p+\psitwos +p$, as a function of the $\psitwos$
  rapidity $y$ for $pp$ centre-of-mass energies of $7$, $8$ and $14$
  TeV. The columns labelled $S^2(W_\pm)$ give the survival factors for
  the two independent subprocesses at different $\gamma p$
  centre-of-mass energies $W_\pm$.} 
\label{tab:A1}
\end{table}

In hadron collisions we also have to take into account additional soft
interactions between the colliding hadrons, which can destroy the
exclusive signature (rapidity gap) of the event. The necessary gap
survival factors for $\psitwos$ production are calculated using the
two-channel eikonal model from~\cite{Khoze:2013dha}. They are
displayed in Table~\ref{tab:A1} for the three different $pp$
centre-of-mass energies of $7$, $8$ and $14$ TeV and for a large range
of rapidities as relevant for the LHC experiments. The columns labelled 
$S^2(W_+)$ and $S^2(W_-)$ give the suppression factors for the two different
$\gamma p$ energies $W_\pm$ as a function of rapidity for each $pp$ 
centre-of-mass energy. 

\begin{figure}
\begin{center}
\includegraphics[width=0.5\textwidth,angle=-90]{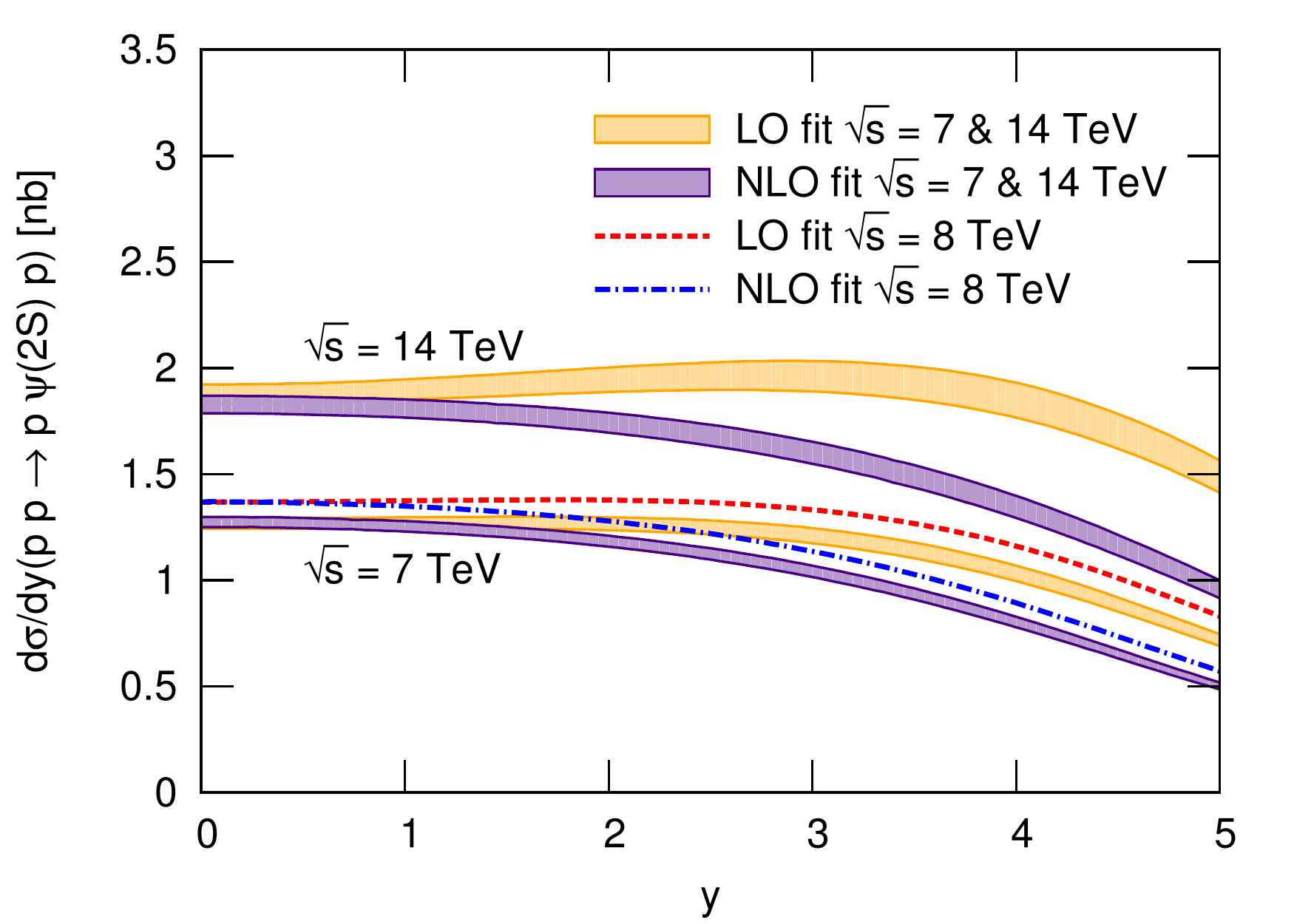}
\caption{Prediction of the exclusive $pp \to p + \psitwos + p$ cross
  section as a function of the $\psitwos$ rapidity $y$ for $pp$
  centre-of-mass energies $\sqrt{s}=7$ and $14$ TeV (shaded bands) and
  $\sqrt{s}=8$ TeV (dashed and dash-dotted lines). 
  The width of the shaded bands indicates only the $1\sigma$
  uncertainty from the $\J$ experimental data used in the gluon fits.
  The uncertainties of the $8$ TeV predictions are very similar to the ones 
  shown for $7$ TeV and are not displayed.} 
\label{fig:psicomb}
\end{center}
\end{figure}

Our theoretical prediction for the exclusive $\psitwos$ production in
ultraperipheral $pp$ collisions, $\mathrm{d}\sigma(pp)/\mathrm{d}y$, in terms of our exclusive
photoproduction cross sections, $\sigma_\pm(\gamma p)$, for the
two subprocesses $\gamma p \to \psitwos\,p$ at energies $W_\pm$ 
is therefore given by
\begin{equation}
\frac{\mathrm{d}\sigma(pp)}{\mathrm{d}y} \ =\  S^2(W_+)\,
\left(k_+ \frac{\mathrm{d}n}{\mathrm{d}k}_+\right)
\sigma_+(\gamma p) \ +\  S^2(W_-)\, \left(k_-
  \frac{\mathrm{d}n}{\mathrm{d}k}_-\right) \sigma_-(\gamma
p)\,. 
\label{eq:sigmappth}
\end{equation}
The photon energies are given by $k_\pm \approx
(M_{\psiprime}/2)\exp(\pm |y|)$ and the photon fluxes are calculated as
described in~\cite{jmrt}. Our cross section predictions are shown in
Fig.~\ref{fig:psicomb} for the three $pp$
centre-of-mass energies of $7, 8$ and $14$ TeV. As in
Fig.~\ref{fig:sigmapsi2sw}, the bands only indicate the experimental
uncertainty of the gluon fit parameters used for the LO and NLO predictions.

In summary, following and supplementing~\cite{jmrt}, we have predicted
the cross section for exclusive $\psitwos$ production in
ultraperipheral $pp$ collisions at the LHC, using gluon
parametrisations extracted from HERA and LHC exclusive $\J$
production data. In principle, once precise $\psitwos$ data become
available, they could be included in a combined analysis, together
with $\J$ and possibly $\Upsilon$ data. Such an analysis,
depending on the accuracy of the data, will require a more detailed 
understanding of the relativistic corrections from the vector meson wave 
functions and would benefit from a complete next-to-leading order 
prediction of the underlying elastic vector meson production
process, $\gamma p \to {\rm V}\, p$, which is beyond the scope of this paper.

\section*{Acknowledgements}

We thank Ronan McNulty for interesting discussions and for encouraging
us to make these predictions. MGR thanks the IPPP at the University of 
Durham for hospitality. This work was supported by the grant 
RFBR 14-02-00004 and by the Federal Program of the Russian State
RSGSS-4801.2012.2.

\end{document}